\documentclass[10pt,conference]{IEEEtran}
\IEEEoverridecommandlockouts
\usepackage[nocompress]{cite}
\usepackage[pdftex]{graphicx}
\graphicspath{{../pdf/}{../jpeg/}}
\DeclareGraphicsExtensions{.pdf,.jpeg,.png}
\usepackage{amsmath}
\usepackage{array}
\usepackage[caption=false,font=footnotesize,labelfont=sf,textfont=sf]{subfig}
\usepackage{url}
\usepackage[T1]{fontenc}
\usepackage[utf8]{inputenc}
\usepackage[english]{babel}
\usepackage{amsmath}
\usepackage{amsfonts}
\usepackage{amssymb}
\usepackage{mathtools}
\usepackage{braket}
\usepackage{amsthm}
\usepackage{pgfgantt}
\theoremstyle{definition}
\newtheorem{theorem}{Theorem}

\newtheorem{remark}{Remark}
\usepackage[inline, shortlabels]{enumitem}
\usepackage{caption}
\usepackage{cellspace}
\usepackage{makecell}
\usepackage{listings}
\usepackage[pdftex=true,hyperindex=true,colorlinks=false,hidelinks=false]{hyperref}
\usepackage{tabularx}
\usepackage[lined,boxed,algonl]{algorithm2e}
\usepackage{etex}
\usepackage{tikz}
\usepackage{pgfplots}
\usepackage{placeins}

\usepackage{cite}
\newenvironment{aeq}{\begin{equation}
	\begin{aligned}
}{
	\end{aligned}
\end{equation}}
\numberwithin{equation}{section}
\numberwithin{theorem}{section}
\numberwithin{lemma}{section}
\numberwithin{definition}{section}

\definecolor{color1}{RGB}{0,0,90} 
\definecolor{color2}{RGB}{0,20,20} 

\usepackage{hyperref} 
\hypersetup{hidelinks,colorlinks,breaklinks=true,urlcolor=color2,citecolor=color1,linkcolor=color1,bookmarksopen=false,pdftitle={Security and Implementation of a scalable quantum-resistant key distribution protocol: Distributed Symmetric Key Exchange},pdfauthor={}}

\usepackage{cleveref}

\crefname{theorem}{Theorem}{Theorems}
\Crefname{theorem}{Theorem}{Theorems}
\crefname{thm}{Theorem}{Theorems}
\Crefname{thm}{Theorem}{Theorems}
\crefname{assump}{Assumption}{Assumptions}
\Crefname{assump}{Assumption}{Assumptions}
\crefname{problem}{Problem}{Problems}
\Crefname{problem}{Problem}{Problems}
\crefname{conjecture}{Conjecture}{Conjectures}
\Crefname{conjecture}{Conjecture}{Conjectures}
\crefname{proposition}{Proposition}{Propositions}
\Crefname{proposition}{Proposition}{Propositions}
\crefname{prop}{Proposition}{Propositions}
\Crefname{prop}{Proposition}{Propositions}
\crefname{cor}{Corollary}{Corollaries}
\Crefname{cor}{Corollary}{Corollaries}
\crefname{lem}{Lemma}{Lemmas}
\Crefname{lem}{Lemma}{Lemmas}
\theoremstyle{definition}
\crefname{definition}{Definition}{Definitions}
\Crefname{definition}{Definition}{Definitions}
\crefname{defn}{definition}{definitions}
\Crefname{defn}{Definition}{Definitions}
\crefname{remark}{Remark}{Remarks}
\Crefname{remark}{Remark}{Remarks}
\crefname{rmk}{Remark}{Remarks}
\Crefname{rmk}{Remark}{Remarks}
\crefname{example}{Example}{Examples}
\Crefname{example}{Example}{Examples}
\crefname{table}{Table}{Tables}
\Crefname{table}{Table}{Tables}
\crefname{align}{}{}
\Crefname{align}{}{}
\crefname{equation}{eq.}{eqs.}
\Crefname{equation}{Eq.}{Eqs.}
\crefname{step}{Step}{Steps}
\Crefname{step}{Step}{Steps}
\crefname{protocol}{protocol}{protocols}
\Crefname{protocol}{Protocol}{Protocols}
\crefname{algorithm}{algorithm}{algorithms}
\Crefname{algorithm}{Algorithm}{Algorithms}
\begin{document}
\title{Distributed Symmetric Key Establishment: \\ a Scalable Quantum-Safe Key Distribution Protocol}

\author{
    \IEEEauthorblockN{
        Jie Lin\IEEEauthorrefmark{1}\IEEEauthorrefmark{2}\IEEEauthorrefmark{10},
        Hoi-Kwong Lo\IEEEauthorrefmark{1}\IEEEauthorrefmark{2}\IEEEauthorrefmark{11},
        Jacob Johannsson\IEEEauthorrefmark{1}\IEEEauthorrefmark{2}\IEEEauthorrefmark{12},
        Mattia Montagna\IEEEauthorrefmark{1}\IEEEauthorrefmark{13},
        Manfred von Willich\IEEEauthorrefmark{1}\IEEEauthorrefmark{14}
    }
    \IEEEauthorblockA{
        \IEEEauthorrefmark{1}Quantum Bridge Technologies Inc., 108 College St., Toronto, ON, Canada
    }
    \IEEEauthorblockA{
        \IEEEauthorrefmark{2}Department of Electrical and Computer Engineering, University of Toronto, 10 King’s College Road, Toronto, ON, Canada 
    }
    \IEEEauthorblockA{
        \{\IEEEauthorrefmark{10}jie.lin, \IEEEauthorrefmark{12}jacob.johannsson, \IEEEauthorrefmark{13}mattia.montagna, \IEEEauthorrefmark{14}manfred.vonwillich\}@quantumbridgetech.com, 
        \IEEEauthorrefmark{11}hklo@ece.utoronto.ca
    }
} 

\maketitle
{
    \begin{abstract}
\label{sec:abstract}

Pre-shared keys (PSK) have been widely used in network security. Nonetheless, existing PSK solutions are not scalable. Moreover, whenever a new user joins a network, PSK requires an existing user to get a new key before they are able to communicate with the new user. The key issue is how to distribute the PSK between different users. Here, we solve this problem by proposing a new protocol called Distributed Symmetric Key Establishment (DSKE)\footnote{DSKE has had several previous names. Firstly, QKI (Quantum Key Infrastructure), then secondly, Distributed Symmetric Key Exchange, before the current name. The first change was due to the name not being descriptive of the protocol, and the second change was because "key establishment" is a more common term in the field than "key exchange".}. DSKE has the advantage of being scalable. Unlike standard public key infrastructure (PKI) which relies on computational assumptions, DSKE provides information-theoretic security in a universally composable security framework. Specifically, we prove the security (correctness and confidentiality) and robustness of this protocol against a computationally unbounded adversary, who additionally may have fully compromised a bounded number of the intermediaries and can eavesdrop on all communication. DSKE also achieves distributed trust through secret sharing. 

We present several implementations of DSKE in real environments, such as providing client services to link encryptors, network encryptors, and mobile phones, as well as the implementation of intermediaries, called Security Hubs, and associated test data as evidence for its versatility. As DSKE is highly scalable in a network setting with no distance limit, it is expected to be a cost-effective quantum-safe cryptographic solution to the network security threat presented by quantum computers. 

\end{abstract}
    \section{Introduction}
\label{sec:introduction}


Pre-shared keys (PSKs) have been used in cryptography for thousands of years \cite{Leighton1969}. A pre-shared key is a secret that is shared between two parties via a secure channel prior to using it to secure communication. PSKs are widely used in commercial applications, such as sensor networks \cite{Chan2003}, banking \cite{ansi2023}, and government \cite{usarmy_keyloader}, as they are an encryption solution with high security and low computational complexity. 

PSK can be provably secure\footnote{Here, we do not refer to the security of the algorithm using the PSK.} when the pre-sharing of keys is secure. In order to have a robust PSK system, keys must have sufficient length and randomness to protect against brute-force attacks, keys should be rotated in regular intervals to mitigate undetected potential compromises of keys, and keys must be securely distributed prior to communication. However, because an individual client is often unable to manually distribute a key to other parties in the network, they need a trusted key distribution centre. From here, key distribution would occur through a secure channel\footnote{Typically, this takes the form of a physical shipment in a PSK key loader.} to each client in the network. The greatest challenge of a PSK system often lies in the distribution of PSKs and preventing the leakage of information by central entities \cite{Chan2004}. In non-centralized networks, keys would need to be pre-shared between every pair of clients in the network, requiring $\frac{N(N-1)}{2}=\mathcal{O}(N^2)$ PSKs for an $N$-user network, which must also be rotated to maintain security. Moreover, whenever a new user joins a network, each existing user would need to first share a key with the new user before before they can communicate, which is generally impractical in large networks. Even if these shipments were made, secure key distribution during the setup phase can be of varying difficulty, depending on client proximity, geography, and local laws.

To solve the pairwise PSK problem, Key Distribution Centers (KDCs) and Distributed Key Distribution Centers (DKDCs) have been proposed. KDC networks work by having a secure connection between each user and the KDC, making for $N$ secure connections, as opposed to $\mathcal{O}(N^2)$. When a user wants to share a key with other users, the KDC would validate the request, then generate and send a session key to each user in a communication group \cite{Blundo_dkdc_2005}. Centralized networks, such as KDC networks suffer from single-points-of-failure, which has two main implications. Firstly, the KDC stores sensitive information that, if leaked, could compromise every key in the network \cite{Chan2004}. Secondly, the KDC is the bottleneck in this network, so if it were to malfunction or be subject to a denial-of-service attack, key distribution would stop \cite{Blundo_dkdc_2005}. The DKDC helps to resolve this problem by distributing the KDC among several entities, where one entity can fail or can be compromised without disrupting secure key distribution. Unfortunately, DKDCs still suffer from several limitations. First, most proposed DKDC systems, such as Refs. \cite{Blundo_dkdc_2005, Daza_DCSKD_2002, Stinson_broadcast_1997}, are just theoretical outlines showing the existence of the protocol without detail on the protocol's real-world construction, and much less, a proof-of-concept to demonstrate practicality and performance. Second, some protocols, such as Ref. \cite{Daza_DCSKD_2002}, make assumptions about the existence of broadcast and secure channels between users and the KDC entities, when these can be difficult to implement in a practical network, especially globally. Third, some DKDC systems use computational security, such as Ref. \cite{Stinson_broadcast_1997}, and might be vulnerable in the future when attackers increase their computing power.

The combination of the preceding factors poses challenges of trust, scalability, and realizability for both non-centralized and centralized PSK solutions. To address these challenges, we propose a new protocol for distributing PSK, called Distributed Symmetric Key Establishment (DSKE). DSKE relies on a secret sharing technique, which allows for distributed trust among a set of semi-trusted entities, alleviating single-point-of-failure risks. The cost of DSKE's client onboarding process is independent of the size of the network, making a PSK system scalable to a worldwide scale. DSKE realizes a DKDC model. We overcome the aforementioned DKDC limitations by 
\begin{enumerate*} [i) ]
\item providing a concrete construction of our protocol in \Cref{sec:dske} and \Cref{app:dske_detailed}, \item performing both authentication and encryption, hence not relying on idealized assumptions such as broadcast channels, \item executing several concrete experiments to measure the performance and integrability of DSKE in real devices (\Cref{sec:implementation}), and \item providing a rigorous proof of information-theoretic security in a composable security framework against a computationally unbounded eavesdropper that can observe all communication, and control under a threshold number of Security Hubs (\Cref{subsec:dske_security}).
\end {enumerate*}


The remainder of this paper is organized as follows. In \Cref{sec:background}, we provide a background to our work. We then discuss related work and our contributions in \Cref{sec:tech_landscape}. In \Cref{subsec:dske_conceptual}, we discuss the details of the DSKE protocol. We then discuss the security of the protocol in \Cref{subsec:dske_security}. We review the current implementations of DSKE in \Cref{sec:implementation}. We analyze the future of DSKE and conclude in \Cref{sec:discussionconclusion}.

    \section{Background}
\label{sec:background}

\subsection{Quantum-Safe Networks}
\label{subsec:quantum_safe}
Quantum computers threaten a cryptography apocalypse due to the famous 1994 Shor's quantum algorithm for efficient factoring \cite{Shor_1997, nistir8105, ETSI_quantum_safe, Mosca2018, threat_quantum_2022, csfc2022, rfc8784}. In particular, for our vast Public Key Infrastructure based on asymmetric cryptography, which secures the majority of the internet today, a transition towards quantum-safe solutions has begun. Innovation can be classified into three main categories: enhancement of Public Key Infrastructure (PKI) referred to as Post-Quantum Cryptography (PQC), Quantum Key Distribution (QKD), and PSK distribution. Due to PKI being highly scalable and deeply ingrained in modern society, there is pressure to develop PQC. However, even with new algorithms, PQC will suffer from being computationally expensive, and not being provably secure against a computationally unbounded adversary. Indeed, several candidate PQC algorithms for the US National Institute of Standards and Technology (NIST) have recently been broken by PCs, thus highlighting PQC's inherent risk, such as SIKE \cite{Townsend_2022} and Rainbow \cite{10.1007/978-3-031-15979-4_16}. If PQC methods are adopted, an unending "cat-and-mouse" game will ensue between code-makers and code-breakers, putting future data at risk \cite{quantum2030}. QKD is a key distribution technique that allows two parties to obtain a shared key for use in symmetric cryptography. QKD is provably secure, and eavesdroppers can be detected during a key exchange \cite{bb84, Xu2020}. However, QKD has limited range, key rate\footnote{The maximum theoretical length of a QKD link that supports relatively high key rates ($\sim$~100 bits/s) is only about 400~km \cite{Lucamarini_Yuan_Dynes_Shields_2018}.} \cite{Lucamarini_Yuan_Dynes_Shields_2018}, and requires dedicated optical fibres and expensive appliances for each communication link\footnote{Quantum repeaters are promising for improving key rate over long distances \cite{Azuma2023}, however, this only increases the cost of each link.}. This, in addition to the requirement of $\mathcal{O}(N^2)$ QKD links in a mesh network, leads to similar scalability issues as with pairwise PSKs.

\subsection{Real-World PSK Use Cases}
\label{subsec:quantum_safe}
A sensor network is a distributed network of small, low-cost devices (called nodes) equipped with sensors that collect and transmit data about the physical environment for various applications \cite{Chan2003, Chan2004}. Depending on the use case, information travelling between nodes can be highly confidential. Numerous methods of node-to-node encryption exist, including asymmetric encryption, a network-wide PSK, and pairwise-shared PSKs \cite{Chan2004}. However, asymmetric encryption is computationally expensive and not provably secure; a network-wide PSK can lead to a network-wide breach if a single node is compromised; while being the most secure option, pairwise-shared keys are difficult to scale in a large network \cite{Chan2004, rfc8784}. 

For banking, the American National Standards Institute (ANSI) for Financial Services' symmetric key management standards (ANSI X9.69 Standard) proposes a protocol for dynamically generated PSKs between clients in a network, called Constructive Key Management (CKM). In this network, a central entity combines two or more client secret key components to construct a key, on-demand, between clients \cite{ansi2023}, similar to previously mentioned KDCs. Dynamically generating keys pushes PSKs forward, as this algorithm reduces the storage need and setup time of the network, although, single-point-of-failure risks arise. 

Regarding government communications, there is also much interest in PSK solutions to encrypt sensitive information \cite{threat_quantum_2022,usarmy_keyloader}, where physical solutions, such as secure key loaders, onto which key material is loaded and shipped physically prior to communication, are used. There are several proposed standards related to PSK in the industry, such as RFC 8696 \cite{rfc8696}, RFC 8784 \cite{rfc8784}, RFC 9257 \cite{rfc9257} and RFC 9258 \cite{rfc9258}.
    \section{Technological Landscape}
\label{sec:tech_landscape}

\subsection{Related Work}
\label{subsec:related_work}
We now discuss similar protocols to DSKE, in the use of secret sharing as a primitive, and a basis on DKDCs, and highlight why our work is very different from those proposals.

We begin by discussing secret sharing based protocols. Ref. \cite{lou_spread_2004} proposed a security protocol called SPREAD for reliable data delivery in a mobile ad hoc network. Even though SPREAD uses the threshold secret sharing scheme to enhance the security, the information-theoretic security was not considered. Indeed, it relied on public key infrastructure to establish secret keys between the sender and the receiver. While it is possible to enhance the security of SPREAD with PSKs, such a combined solution will encounter the scalability issue of PSK. On the other hand, DSKE is a scalable and provably information-theoretically secure protocol.

Ref. \cite{nishimura_reinforcement_2010} used the threshold secret sharing scheme and a multipath routing technique to enhance the security of vocal communication over an open network. Unlike our work, which endorses information-theoretical security, this paper uses encryption based on computational assumptions. Their use of the secret sharing scheme is also a straightforward application without considering how to validate the correctness of the reconstructed secret. In DSKE, our novel secret-authenticating tag guarantees the correctness of the reconstructed secret. 

We now shift to discussing DKDC based protocols. Ref. \cite{Blundo_dkdc_2005} (and the same authors' earlier conference abstract \cite{Blundo_abstract_2000}) presents an information-theoretically secure abstract DKDC system. A secure channel is assumed between all servers and between each server and each client without specifying how to realize such a channel. Each time when a key request between two clients are made, secure channels need to be invoked. However, in DSKE, we only require secure channels during the initial set-up phase, and we discuss how to realize those channels concretely. Similar to DSKE, a secret sharing scheme\footnote{For the theoretical protocol, the specific scheme does not matter, Shamir's secret sharing scheme would fit security requirements for the network.} is employed to mitigate both Denial-of-Service attacks and direct attacks. Moreover, Ref. \cite{Blundo_dkdc_2005} considers a general group key, which is beyond the scope of our paper.

Ref. \cite{Daza_DCSKD_2002} aims for a more realistic implementation of a DKDC, requiring only authenticated channels between the users and the servers, rather than secure channels. Similar to Ref. \cite{Blundo_dkdc_2005}, a group key is computed between any subset of users in the network. It also uses a secret sharing scheme and encrypts the share using a homomorphic encryption scheme. The major difference between this protocol and DSKE is that Ref. \cite{Daza_DCSKD_2002} assumes a computationally bounded adversary, whereas DSKE is information-theoretically secure, using secure channels during the initial setup phase.


Ref. \cite{Stinson_broadcast_1997}, describes information-theoretically secure methods for secure key predistribution and broadcast encryption among a network of users and a Trusted Authority (TA). This differs from DSKE in multiple ways. DSKE does not use a broadcast channel. DSKE uses a distributed central entity instead of a TA, and these entities in DSKE do not initiate key distribution, rather they re-encrypt and relay key information.


\subsection{Our Contribution}
\label{subsec:our_contribution}

DSKE is a novel protocol that can provide the following features simultaneously:

\begin{itemize}[leftmargin=1em]
    \item Information-theoretic security. It guarantees confidentiality and correctness.
    \item Scalability I: low cost to add a new user. The cost of onboarding each new client is proportional to the number of Security Hubs rather than to the number of clients. 
    \item Scalability II: no extra burden to existing users when onboarding a new user. When a new user joins the network, there is no need for the existing user to receive securely distributed key material from a trusted key distribution centre before it can communicate with the new user.
    \item Distributed trust. The DSKE network removes a single point of failure. To compromise the security of DSKE, an adversary needs to compromise enough (independent in principle) security entities called Security Hubs.
    \item Unlimited distance. There is no fundamental communication distance limit, unlike QKD solutions.
    \item No extra pre-shared keys needed for message authentication. DSKE allows the co-transmission of the key and message, while a typical message authentication scheme requires a pre-shared key to securely verify the message’s authenticity.
    \item Software-based solution. Similarly to PQC, DSKE does not require dedicated hardware. This makes the adoption of DSKE much more cost-effective than QKD solutions. 
\end{itemize}

To our best knowledge, no other known protocol can achieve the same features simultaneously.
 
We emphasize that the use of threshold secret sharing scheme in DSKE is not a trivial application. DSKE uses multiple parallel executions of the Shamir's secret sharing scheme and has a novel design of secret-authenticating tag that allows the validation of secret without pre-shared keys between the sender and the receiver\footnote{Many existing key distribution schemes in the literature assumes some other means of authentication, typically in a form of pre-shared keys between the sender and the receiver.}. Pre-shared random data are shared only between each user and each Security Hub, with far fewer Hubs than users, making DSKE more scalable than other PSK solutions. 


    \section{The DSKE Protocol}
\label{sec:dske}

{
    \subsection{Conceptual Outline}
\label{subsec:dske_conceptual}
DSKE makes PSK solutions scalable in a large network. Here, we give a synopsis of the DSKE protocol, which was first proposed in \cite{Lo2022}, while also referencing the more detailed description of the protocol described in \Cref{app:dske_detailed}.

\begin{figure}[t]
    \includegraphics[width=0.9\linewidth]{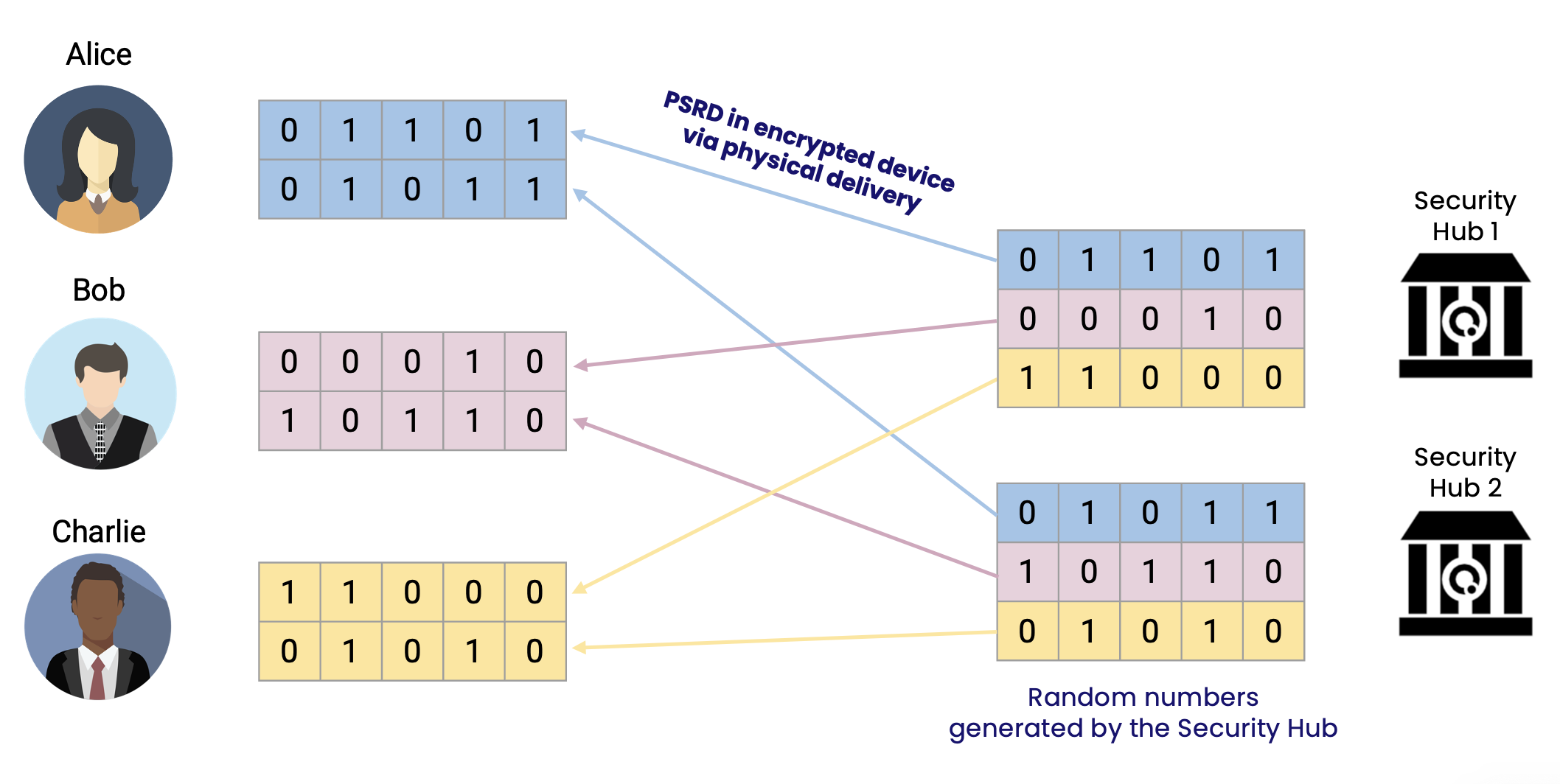}
    \centering
    \caption{The results of the one-time set-up: Steps 1 (\textit{PSRD generation and distribution}) and 2 (\textit{Peer identity establishment}) of the protocol. DSKE users Alice, Bob and Charlie share an ordered table of PSRD with each of the Security Hubs. Each Security Hub only knows its own part of the users' tables. In this illustration, the PSRD is shown as bits.}
    \label{figs:protocol}
\end{figure}

We work with a two-user key agreement protocol in a network setting with a large number, $N$, of potential ``end users'' in the presence of a number, say $n$, of third parties called Security Hubs. The Security Hubs behave similar to DKDCs, with some differences that will become apparent. The Security Hubs are numbered from $1$ to $n$ and an identifier $P_i$ is assigned to the $i$th Hub. The main goals are to guarantee that both users agree on the same secret and to protect the privacy of the agreed secret from potential adversaries, including other end users and the Security Hubs. During the one-time set-up (Steps (1) and (2) of the protocol as described in \Cref{app:dske_detailed}), secure channels are assumed between the end users and Security Hubs. Those secure channels enable the end users and the Security Hubs to share some pre-shared random data (PSRD). Once the one-time setup is complete, we are in the situation described in \Cref{figs:protocol}. \Cref{figs:protocol} shows an example of a network with the users such as Alice, Bob, and Charlie together with two Security Hubs. Each user shares a table of PSRD with each Security Hub. Note that each Security Hub knows only the values of its own PSRD, but has no information on the values of the PSRD of other Security Hubs.


For the simplicity of the discussion, we assume one-way communication from Alice to Bob, that is, Alice requests via the Security Hubs to exchange a secret with Bob (This begins Step (3) of the protocol as described in \Cref{app:dske_detailed}). Alice is not interested in receiving information from Bob at all during the execution of the protocol, and may have only unidirectional communication available. Two-way communication can be realized by two separately managed keys, generated by two iterations of the DSKE protocol.


In the DSKE protocol, Alice generates $n$ shares using PSRD shared between Alice and each Security Hub in an $(n, k)$-threshold scheme\footnote{For technical reasons, multiple parallel executions of a chosen $(n, k)$-threshold scheme are needed, where the number of executions depends on the message length.} of Shamir's secret sharing scheme \cite{shamir1979share}, where $k$ is the minimum number of shares needed to reconstruct the secret. She also generates a secret-authenticating tag $o^A:=h'_{u^A}(S^A)$, where $u^A \parallel S^A$ is the secret from the $(n, k)$-threshold scheme, and $h'_{u^A}$ is a hash function with its parameter $u^A$, which is chosen from a family of 2-universal hash functions. She encrypts each share through one-time pad (OTP) with PSRD, then sends the $i$th share $Y_i$ and the secret-authenticating tag to the Hub $P_i$ via authenticated channels. We note that each Hub's secret-authenticating tag is the same. 

After receiving the secret-authenticating tag and the encrypted share, an honest Hub decrypts the share, and then re-encrypts the share using the PSRD shared between the Hub and Bob. It forwards the secret-authenticating tag and the newly encrypted share to Bob via an authenticated channel. 

After Bob receives enough messages from Hubs, he reconstructs a candidate value of the secret from each subset of $k$ of the shares received. Then, in the secret validation step (Step (4) in \Cref{app:dske_detailed}), he validates each possible candidate secret value against the secret-authenticating tag, which is chosen to be the same tag sent by at least $k$ Hubs. If there is no secret that passes the secret validation step, he aborts the protocol.




    \subsection{Security of DSKE}
\label{subsec:dske_security}

Normally, a message authentication code would be employed to allow detection of such any change in the reconstructed secret, but this needs a shared key to implement.  Transmitting a validation key via the same secret sharing scheme violates the normal premise for authentication: that the validation key is assured to be the same at both sides. We remark that Ref. \cite{Daza_DCSKD_2002} assumed authenticated channels, and Ref. \cite{Stinson_broadcast_1997} assumed broadcast channels. In contrast, DSKE requires neither. In our work, our construction of secret-authenticating tag allows one to transmit such a key using the same secret sharing scheme while providing authenticity, under the same premise that the secret sharing scheme already has. 

We briefly discuss the security of DSKE and leave technical details to a separate paper \cite{lin2023composable}.  First, we discuss the relevant threat model. A collection of adversarial entities can include a coalition of end users other than Alice and Bob, eavesdroppers, and a subset of the Security Hubs. A compromised Security Hub may deviate from the protocol. No limits are placed on compromised hubs, but they cannot access or modify any confidential information held by other parties. This set of adversaries may collude to attempt to compromise the objective of the protocol between Alice and Bob. We call this collection of adversarial entities \textit{Eve}. As a robustness analysis of a protocol is concerned with an honest implementation of the protocol, which is a modified threat model from that of the correctness and confidentiality analysis, we call Eve's behaviour \textit{passive Eve}: Eve is passive on all communication links, that is, she is allowed to listen to all the communications except for the initial sharing of tables by honest Security Hubs but she does not tamper; she is still given the ability to fully control compromised Security Hubs. When Eve is passive, we show that the DSKE protocol completes (i.e., does not abort) with a high probability.

We then list assumptions used in our security proof:
\begin{enumerate}[label = \roman*),leftmargin=1em]
    \item The pre-shared random data (PSRD) are securely delivered by all uncompromised Security Hubs to both Alice and Bob. By \textit{securely delivered}, we mean ensured confidentiality, integrity and proper identity verification through secure channels between Security Hubs and clients\footnote{PSRD can be delivered by physically shipping a secure data storage device or via QKD links.}.
    \item The two users, Alice and Bob, are both honest.
    \item A number of the Security Hubs might be compromised, and this number has a known upper bound. 
    \item For the robustness analysis, a number of the involved Security Hubs might malfunction, either due to unavailability, communication failure\footnote{This type of malfunction includes the a communication link providing an incorrect sender identity to the receiver.}, or compromise, and this number has a known upper bound. This is incorporated with the assumption that Eve is passive on the communication links.
\end{enumerate}

As a cryptographic protocol is often combined with many other protocols, it is important to prove the security of the protocol in a composable security framework \cite{Maurer2011, Portmann2022}. The composability result in such a framework asserts that in analyzing the security of a complex protocol, one can simply decompose it into various subprotocols and analyze the security of each. Provided that each real subsystem constructed by a subprotocol is close to an ideal subsystem within some $\epsilon$, which is quantified by some distance measure, the real system constructed from the combined protocol will then be close to the combined ideal system. The sum of the $\epsilon$-values for the subprotocols gives an $\epsilon$-value for the combined protocol. We analyze the security of the DSKE protocol in the framework of constructive cryptography (see \cite{Maurer2011, Portmann2022} for further details about the framework). We state the main theorems here and leave proofs to a separate paper, \cite{lin2023composable}.

\begin{theorem}[Security of the DSKE protocol]\label{thm:security_general}
    ~\\The DSKE protocol using an $(n, k)$-threshold scheme is $\epsilon + 2n \epsilon'$-secure, where $\epsilon = \min({n \choose k}\frac{m+1}{|F|}, 1)$ and $\epsilon' = \min(\frac{s}{|F|},1)$. Here, $|F|$ is the number of elements of the field $F$, $m$ is the number of field elements in the final secret, and $s$ is the number of field elements in the authenticated message. 
\end{theorem}

\begin{theorem}[Robustness of the DSKE protocol]\label{thm:robustness_general}
    When the upper bound on the number of compromised Security Hubs is no greater than $\min(n-k, k-1)$, the DSKE protocol using an $(n, k)$-threshold scheme  is $\epsilon$-robust with $\epsilon = \min({n \choose k} \frac{m+1}{|F|}, 1)$,  where $|F|$ is the number of elements of the field $F$, and $m$ is the number of field elements in the final secret.
\end{theorem}

}

    \section{Implementations of DSKE}
\label{sec:implementation}


We now examine several Proof-of-Concept implementations of DSKE. Demonstrating the Proof-of-Concept for DSKE is important at this stage because it tests the theoretical scalability and security propositions of DSKE and provides a preview of its potential extensive utilization in the future.

\subsection{Share Performance}
For this demonstration, we used Shamir's secret sharing scheme over GF$(2^8)$, with two clients and multiple Security Hubs communicating over TCP/IP with only the unidirectional messages. This implementation allowed us to run performance tests on the system. In particular, we demonstrate performance for a range of $n$ and $k$ (where $1\leq k \leq n$) for the number of shares $s$ received by Bob in the range $k \leq s \leq n$, which means that zero or more of the $n$ shares are corrupted, either with a mismatched tag (as for a data corruption in transit) or with a matching tag (as for a compromised Security Hub attacking the system). The agreed secret length was set to 8 megabits, which is large enough to show the asymptotic scaling behavior of the processing rate. We implemented the receiver validation for the quickest operation when the system is not under active attack, namely by reconstructing the secret from only $k$ shares and then validating the result against the secret-authenticating tag, only proceeding to another combination of shares if this fails. No attempt was made at full optimization.


\begin{figure}[t]

	\includegraphics[width=0.9\linewidth]{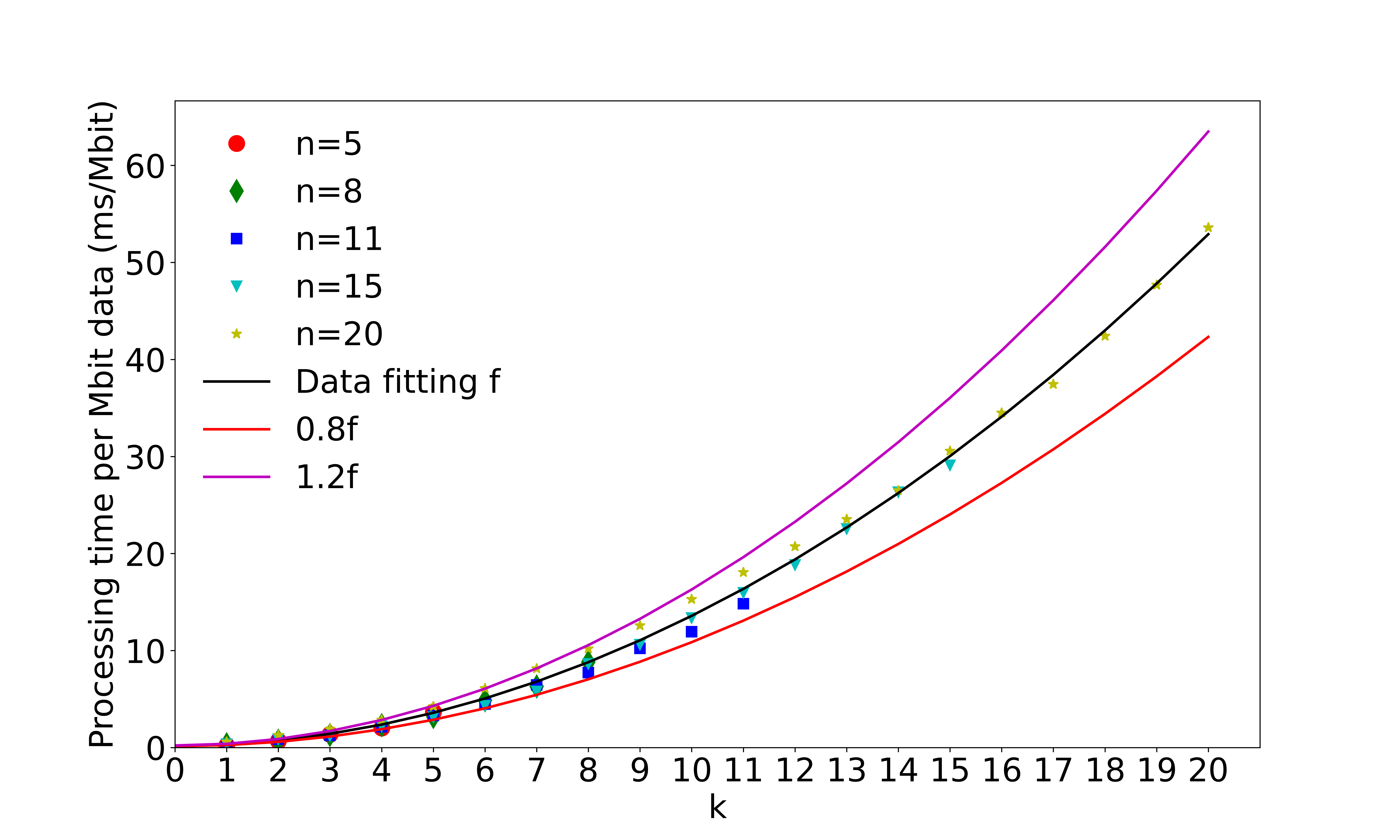}
        \centering
	\caption{Scaling behavior of processing time as a function of $k$ for different values of $n$ with $k \leq n$. The data fitting curve is $f(k) = 0.141 k^{1.977} + 0.183$. All data points fall in between $0.8f(k)$ and $1.12f(k)$. The processing time scales close to proportionally with $k^2$ with minimal dependence on $n$.}
	\label{fig:run_time}
\end{figure}

Processing cost for Alice and Bob as affected by scaling of $n$ and $k$ is shown in \Cref{fig:run_time}. For this analysis, we omitted communication and data retrieval costs, which can be separately estimated and will scale approximately linear with the amount of data communicated by the protocol, including the message cryptographic primitives. The results here inherently reflect our design choices. Here, secret reconstruction included first deriving the polynomial coefficients, and then using these to derive the secret and the other shares as needed. The step of deriving the shares from the coefficients was found to take nearly negligible time compared to the generation of the coefficients. These results demonstrate an implementation where the processing cost at each of $A$ and $B$ scales close to proportionally with $k^2$ with minimal dependence on $n$; this is sufficient to show that implementations for realistic values of $n$ and $k$ can be performant. Example parameter values might be $n=9$ and $k=5$, with share processing time in the order of $1$ ms/Mbit at each client for our implementation. As the threshold value $k$ is supposed to be a small constant, e.g. 3 to 5, the DSKE protocol can be efficiently implemented. 

\subsection{Layer 3 VPN \& QKD Experiment}
\label{subsec:wireguard_qkd}

In larger DSKE networks, the PSRD delivery from Hub to client using a secure channel becomes increasingly difficult. For this reason, Security Hubs can be split in two types of entities, multiple Local Distributors and the Security Server. The collection of Local Distributors takes over the responsibility of PSRD distribution and generation from the central Security Server, which performs all other Security Hub functions. One can imagine that a Security Hub would set up Local Distributors in areas of relatively high client density to shorten PSRD delivery routes. Because the Local Distributor is generating the PSRD, it would also need to send a copy to the Security Server. However, it can send arbitrarily large blocks of PSRD, limiting the number of deliveries.

This experiment showcases how DSKE can be combined with traditional asymmetric key algorithms, adding an extra layer of symmetric encryption for significantly increased security. The DSKE network was set up with 2 clients, and 2 Security Hubs, each with 2 Local Distributors. These 2 clients were externally communicating via a WireGuard\textsuperscript{\tiny\textregistered}\footnote{WireGuard is a registered trademark of Jason A. Donenfeld.} Virtual Private Network (VPN) tunnel. WireGuard uses PKI to exchange a session key to encrypt and authenticate communication, and has the option for a supplied PSK to increase security. Additionally, the  experiment was meant to demonstrate the compatibility of a client refilling PSRD with a QKD link between itself and the Local Distributor, allowing PSRD to be sent over a public channel with information-theoretic security.


\subsubsection{Layer 3 VPN}
\label{subsubsec:wireguard}

As previously mentioned, the 2 clients were communicating via a layer 3 WireGuard VPN tunnel. To send data packets between clients, WireGuard employs ChaCha20-Poly1305, an Authenticated Encryption with Associated Data (AEAD) algorithm \cite{wireguard}. It consists of the ChaCha20 stream cipher algorithm, with Poly1305 authentication \cite{rfc8439}. ChaCha20 is similar to OTP encryption, except that the keystream is derived from a counter, 96-bit nonce, and 256-bit key \cite{chacha}. The 256-bit key is generated with the asymmetric Curve25519 ECDH (Elliptic Curve Diffie-Hellman) function \cite{10.1007/11745853_14, 1055638, wireguard}. However, asymmetric algorithms are under threat from quantum algorithms such as Shor's \cite{Shor_1997} and Discrete Log \cite{proos2004shors} algorithms. In order to increase WireGuard security, it allows for PSK injection when deriving the 256-bit key used in ChaCha20 \cite{wireguard}. This experiment was meant to demonstrate the performance of WireGuard VPN with DSKE PSKs, by testing it with and without a DSKE-supplied PSK. Both trials consisted of a 60 second ping and bitrate test. As seen in \Cref{table:wireguard_performance}, performance is essentially equal in both trials.


\begin{table}[ht]
    \renewcommand{\arraystretch}{1.3}
    \caption{WireGuard Communication Statistics Between 2 DSKE Clients}
    \label{table:wireguard_performance}
    \centering
    \begin{tabular}{|c|c|c|}
        \hline
        \textbf{Type of Encryption} & \textbf{Ping (ms)} & \textbf{Bitrate (Mb/s)} \\
        \hline
        Standard WireGuard & $1.608 \pm 0.038$ & $316.433 \pm 0.892$ \\
        \hline
        Standard WireGuard + PSK & $1.679 \pm 0.030$ & $316.683 \pm 0.926$ \\
        \hline
    \end{tabular}
\end{table}

\subsubsection{QKD}
\label{subsubsec:qkd}

Traditionally within DSKE, a client must refill PSRD via physical shipments sent from Security Hub's Local Distributors, delivered by trusted couriers to maintain information-theoretic security. However, QKD also offers information-theoretic security, and can sustain sufficient key rates over shorter distances. In an ideal system, any client should be sufficiently close to a Security Hub's nearest Local Distributor to make a QKD link feasible. This experiment has 4 Local Distributor-client connections. To test the feasibility of using a QKD system, one of the physical (USB drive) delivery chains were replaced by a QKD link, allowing PSRD to be sent over a public channel, encrypted using a OTP with QKD keys. The QKD system used was ID Quantique\textsuperscript{\tiny\textregistered}'s ID3200 Clavis$^3$. The two QKD machines were connected by a 200~m long G657A SM-3.0 single-mode optical fiber. 

The experiment tested whether or not QKD was compatible within a larger DSKE network. We observed a quantum key generation rate (and therefore a PSRD delivery rate) of 2800~bits/second. This would be enough to generate $\sim\frac{5}{n}$ DSKE PSKs per second, where $n$ is the number of involved Security Hubs, and which is satisfactory considering that it is common practice to rotate keys in intervals on the order of days \cite{Google_2023}. Beyond the experiment, the Clavis$^3$ system claims to maintain a 1400~bits/second key rate when the optical channel transmission loss is 12~dB \cite{Hernandez_2020}. Using the in-lab optical fiber's attenuation at 1550~nm of 0.25~dB/km\cite{guide_2012} would allow for a 48~km link between Alice and Bob, while maintaining a 1400 bits/second secret key rate, which is good for DSKE purposes.

\subsection{DSKE Device Feasibility \& Performance Experiment}
\label{subsec:isc}

To test multiple DSKE devices, as well as the theoretical propositions of DSKE, a DSKE network was created consisting of 3 Security Hubs, each with a Local Distributor, and 9 clients, 4 of which being DSKE key management entities (KMEs), and 5 being mobile phones with a proof-of-concept DSKE messaging application. The DSKE network was physically set up in two colocations, Montreal and Ottawa. The 3 Security Hubs, 2 located in AWS and one located in an Ottawa colocation, each consisted of 1 Local Distributor installed on a standard laptop, from which PSRD would be sent. ID Quantique's QUANTIS-USB-4M Quantum Random Number Generator (QRNG), with 4 Mbit/s data transfer rate, was plugged into the laptops to generate PSRD, as the security proof (\Cref{subsec:dske_security}) assumes truly random PSRD.

\subsubsection{Key Management Entity}
\label{subsubsec:kme}

The KME provides DSKE client services that can be installed to provide PSK in existing infrastructure. The KME is wrapped in a standardized API, the ETSI QKD GS 014, allowing the KME to interface with a large variety of network appliances\cite{etsiqkd2019}. In this experiment, 2 KMEs were set up with commercial firewall devices and 2 with commercial link encryptors. Initial PSRD delivery occurred via a secure physical drive. The 2 firewall appliances were set up in 2 colocations, one in Montreal and one in Ottawa, with an IPsec based layer 3 ---networking layer--- VPN tunnel protected by DSKE PSKs. In the same network, the 2 link encryptors were set up again in the same 2 colocations, with a layer 2 ---data link layer--- tunnel protected by DSKE PSKs. DSKE's ability to interface at both link and network levels is important, since they present different use cases. Link tunnels are good at secure data transfer, but are difficult to scale, while network tunnels are good at scaling and long-distance data transfer, but have higher latency and reduced performance.

\subsubsection{DSKE Messaging App with QR Code Implementation}
\label{subsubsec:black_phone}

The DSKE messaging app acts as a client within the DSKE network. It is a proof-of-concept iOS\textsuperscript{\tiny\textregistered}\footnote{iOS is a registered trademark of Cisco.} application installed on 5 iPhone XRs\textsuperscript{\tiny\textregistered}, configurable for text chat and voice/video calls. Voice/video calls were made using webRTC\textsuperscript{\tiny\textregistered} and chatting was done using a broker service, Metered\textsuperscript{\tiny\textregistered}, with end-to-end encryption and authentication through DSKE PSKs. Initial PSRD delivery from the Security Hubs occurred by scanning of QR codes that were generated and sent online by the Local Distributors. This app was used to demonstrate the flexibility of DSKE if the initial PSRD delivery was truly secure. 

\subsubsection{Tests Results}
\label{subsubsec:results}

Overall, the results show that the theoretical propositions of DSKE are applicable in real networks. 

\begin{enumerate}[leftmargin=1em]
    \item \textbf{Information-Theoretic Security}: A security consulting third-party verified the composable security proof of DSKE.
    \item \textbf{Distance Scalability}: The use of mobile devices, connected via a public internet connection, implies distance scalability.
    \item \textbf{Size Scalability}: While onboarding clients, the process remained identical, independent of the network size.
    \item \textbf{Fault tolerance and no single point of failure}: Clients could exchange keys only with $\ge$ 2 Hubs online\footnote{It is important that they shouldn't be able to communicate with 1 Hub, since the single point of failure problem resurfaces.}.
    \item \textbf{Black box pen test}: The KME and DSKE passed a third-party company's black box penetration test suite.
    \item \textbf{Performance}: 2 virtual cloud-operated DSKE clients achieved a maximum key rate in excess of 20 Mbit/s.
\end{enumerate}
    \section{Discussion and Conclusion}
\label{sec:discussionconclusion}

As mentioned in \Cref{subsec:wireguard_qkd}, Local Distributors augment the Security Hub by offloading PSRD distribution duties for a certain geographical region, reducing the distance of client deliveries, allowing DSKE to be implemented in a global setting. Additionally, reducing the client delivery distance allows for the possibility of QKD links, erasing the need for physical shipments altogether (\Cref{subsec:wireguard_qkd}). PSRD generation would occur within the Local Distributor and be shipped in a massive block to the Security Server, given that there are commercial products to allow large-scale, tamper-proof physical delivery of data, for example, \cite{aws_snowball}. Fortunately, high-speed (e.g. 40Mbit/s and 240 Mbit/s) QRNGS are commercially available \cite{ID_Quantique_2024b}, for the generation of this PSRD. Upon making deliveries to clients, the Local Distributor relays the details of the assigned block to the Security Server and deletes it. 

In the future, we can imagine certain institutions providing DSKE Security Hubs. The use of third-party Security Hubs adds to Hub count, increasing decentralization and increasing flexibility when choosing $n$ and $k$. Lastly, it lessens single-point-of-failure risks as different hubs are managed by different entities, differentiating their potential attack schemes.


In summary, the DSKE protocol provides information-theoretically secure key exchange between two honest parties without the need for prior contact. Traditional PSK systems face issues in scalability, as the onboarding cost of a single client grows with the number of clients in the network, and in single points of failure as a KDC holds enough information to determine all PSKs in the network. DSKE solves the onboarding problem by dynamically generating PSKs between clients, molded from pre-shard random data, which is provided without involvement from other clients. Additionally, since trust is distributed among several Security Hubs, DSKE is resilient to hub seizure and denial-of-service attacks.

We have proved (in \cite{lin2023composable}) the security and robustness of the protocol against any computationally unbounded adversary who, in addition to eavesdropping on all communication, may compromise or disable a bounded number of Security Hubs. Through experiments, we see that DSKE is capable of integrating with a wide range of devices, including providing client services to link encryptors, network encryptors, and mobile phones as well as Security Hubs in different environments. In the future, DSKE can function on a global scale by extending a Security Hub to include a fleet of Local Distributors, and in using QKD technology to make the delivery of PSRD on-demand and autonomous. DSKE stands as a scalable and dependable PSK solution, making it an exceptional choice for safeguarding data in the quantum-safe internet.
    \section*{Acknowledgments}
\label{sec:acknowledgements}

The research reported in this paper was supported by the Connaught Innovation Award, the Borealis AI Graduate Fellowship, Mitacs Accelerate, Innovative Solutions Canada (ISC), and Defense Research and Development, Canada (DRDC). H.-K. Lo is supported by NSERC, NRC CSTIP program and CFI.

}

\appendices
\crefalias{section}{appendix}
{
    \section{Detailed Description of DSKE}
\label{app:dske_detailed}

This description of DSKE is adapted from Ref. \cite{lin2023composable}.

\subsection{Parameter choices}
\label{subapp:dske_detailed_parameter}

\begin{itemize}[leftmargin=1em]
    \setlength\itemsep{0.25em}

    \item Determine a finite field $F$, composed of \textit{elements}.
 
    \item Determine the final secret length $m$. 
 
    \item Select the parameters $n$ and $k$ of the sharing scheme. 

 
    \item Hash functions:
    $\{h_{c,d}: (y_{1},\dots,y_{m}) \mapsto d + \sum_{j=1}^m c^{j} y_{j} \}$ and
    $\{h'_{c,d,e}: (y_{(1)},\dots,y_{(m)}) \mapsto d + ce + \sum_{j=1}^m c^{j+1} y_{(j)} \}$.
 
    \item $(n, k)$-scheme: $f : x \mapsto c_{0} + c_{1}x + \cdots + c_{k-1}x^{k-1}$.

    \item Determine an injective mapping $\{0, \dots, n\} \to F : i \mapsto x_i$.

    \item Determine a bijective mapping $g:\{0,\dots,|F|-1\} \to F$.

    \item A Hub sends each client two tables, e.g. $H^{A}_i$ and $\overline{H}^{A}_{i}$.

    \item Validate mutual identifiers $P_{i}, A_{i}, B_i$.

    \item Write $A$ for $A_{i}$, chosen equal for all $i$, as for $B$. \footnote{In a practical system, the clients may track these independently by Hub.}

    \item A message tag validation key is used once only.

\end{itemize}

\subsection{Baseline protocol}
\label{subapp:dske_detailed_baseline}

\begin{enumerate}[label={(\arabic*)}, ref=\arabic*, leftmargin=0.5cm]

    \item \textbf{PSRD generation and distribution}
	
    Hubs securely provide $H^A_i$ and $\overline{H}^{A}_{i}$ to Alice, etc. 
    Alice uses offset $j^A_i$ to track use of $H^A_i$, initialized $j^A_i:=0$. A Hub similarly uses $\overline{j}^A_i$ for $\overline{H}^{A}_i$.  A receiver tracks use of each table element.

    \item \textbf{Peer identity establishment}
    
    Alice and Bob establish each other's identifiers.


    \item \textbf{Secret agreement}

    \begin{enumerate}[label=(\alph*),leftmargin=0.25em, ref=\theenumi{}.\alph*]
        \item \textit{Share generation}
        \begin{enumerate}[label=(\roman*),leftmargin=0.25em,  ref=(\theenumii{}.\roman*)]
            \item Alice retrieves and erases $R^A_i$ (length $3+m$) and $v^A_i$ (length $2$) at offset $j^A_i$ in $H^A_i$, using $j^A_i+5+m$ as $j^A_i$ on the next protocol iteration. \label[step]{step:3a1}

            \item Alice sets:%
            \begin{aeq}
                Y^A_i := R^A_i \quad  \forall i \in \{1,\dots,k\}
            \end{aeq}%
            \begin{aeq}
                f_{-2}(x_i)\parallel\dots\parallel f_{m}(x_i):=Y^A_i ~\forall i \in \{1, \dots, k\}
            \end{aeq}%

            \item Alice solves for $Y^A_i$, using $3+m$ sharing schemes:%
            \begin{aeq}\label{eq:share}
                f_{-2}(x_i)\parallel\dots\parallel f_{m}(x_i)=:Y^A_i ~\forall i \in \{k+1, \dots, n\}
            \end{aeq}%
        \end{enumerate}

        \item \textit{Share distribution}
        \begin{enumerate}[label=(\roman*),leftmargin=0.25em  ]
            \item\label{step:share_distribution_alice} Operations by Alice for share distribution:
            \begin{enumerate}[label=(\arabic*),leftmargin=0.25em ]
                \item Alice solves for $Y^A_0=f_{-2}(x_0)\parallel\dots\parallel f_{m}(x_0)$.

                \item Alice partitions $Y^A_0$ into $u^A$ and $S^A$:%
                \begin{aeq}
                    Y^A_0 =: u^A \parallel S^A
                \end{aeq}%

                \item Alice calculates the \textit{secret-authenticating tag} $o^A$ as%
                \begin{aeq}
                    o^A := h'_{u^A}(S^A)
                \end{aeq}%

                \item Alice calculates%
                \begin{aeq}
                    Z^A_i := Y^A_i - R^A_i \quad i \in \{1, \dots, n\}
                \end{aeq}%
                Note:  $Z^A_i$ is zero for $i \in \{1, \dots, k\}$ due to cancellation.

                \item Alice chooses $K^A$ to make $(A,K^A)$ unique, to get%
                \begin{aeq}
                    M^A_i := A \parallel B \parallel K^A \parallel g(j^A_i) \parallel Z^A_i \parallel o^A
                \end{aeq}%

                \item Alice calculates the message tag $t^A_i$ as%
                \begin{aeq}
                    t^A_i := h_{v^A_i} ( M^A_i )
                \end{aeq}%

                \item Alice sends to Hub $P_i$ for $i \in \{1, \dots, n\}$:%
                \begin{aeq}
                    M^A_i \parallel t^A_i 	
                \end{aeq}%
            \end{enumerate}

            \item Operations by each Hub $P_i$, related to Alice:
            \begin{enumerate}[label=(\arabic*),leftmargin=0.25em ]
                \item Hub $P_i$ splits the sequence from Alice via%
                \begin{aeq}
                    M^A_i \parallel t^A_i
                \end{aeq}%

                \item The Hub splits $M^A_i$ into its components via%
                \begin{aeq}
                    M^A_i =: A \parallel B \parallel K^A \parallel g(j^A_i) \parallel Z^A_i \parallel o^A
                \end{aeq}%

                \item The Hub discards the message if $(P_i, A, B)$ is disallowed, was not received via the routing from $A$, or the $3+m+2$ elements at offset $j^A_i$ in its table $H^A_i$ were used. Discarding here does not deplete table elements.
    
                \item The Hub retrieves and erases $R^A_i$ (length $3 + m$) at offset $j^A_i$ and $v^A_i$ (length 2) at offset $j^A_i + 3 + m$ from the table.

                \item The Hub discards the message on failure of the relation%
                \begin{aeq}
                    t^A_i = h_{v^A_i}( M^A_i ) 	
                \end{aeq}%

                \item The Hub calculates%
                \begin{aeq}
                    Y^A_i := Z^A_i + R^A_i
                \end{aeq}%
            \end{enumerate}

            \item Operations by each Hub $P_i$, related to Bob:
            \begin{enumerate}[label=(\arabic*),leftmargin=0.25em  ]
                \item The Hub retrieves and erases $\overline{R}^B_i$ and $\overline{v}^B_i$ from $\overline{H}^B_i$ using $\overline{j}^B_i$, similarly to \ref{step:3a1}.

                \item The Hub calculates%
                \begin{aeq}
                    \overline{Z}^B_i := Y^A_i - \overline{R}^B_i
                \end{aeq}%

                \item The Hub generates the message $\overline{M}^B_i$:%
                \begin{aeq}
                    \overline{M}^B_i := A \parallel B \parallel K^A \parallel g(\overline{j}^B_i) \parallel \overline{Z}^B_i \parallel o^A
                \end{aeq}%

                \item The Hub calculates the message tag $\overline{t}^B_i$ as%
                \begin{aeq}
                    \overline{t}^B_i := h_{\overline{v}^B_i} (\overline{M}^B_i) 
                \end{aeq}%

                \item The Hub sends to Bob the element sequence%
                \begin{aeq}
                    \overline{M}^B_i \parallel \overline{t}^B_i
                \end{aeq}%
            \end{enumerate}

            \item Operations by Bob for each Hub $P_i$, related to Alice:
            \begin{enumerate}[label=(\arabic*),leftmargin=0.25em ]
                \item Bob splits the sequence from Hub $P_i$ as%
                \begin{aeq}
                    \overline{M}^B_i \parallel \overline{t}^B_i
                \end{aeq}%
    
                \item Bob then splits $\overline{M}^B_i$ into its components%
                \begin{aeq}
                    \overline{M}^B_i =: A \parallel B \parallel K^A \parallel g(\overline{j}^B_i) \parallel \overline{Z}^B_i \parallel o^A
                \end{aeq}%

                \item Bob discards the message if $(P_i, A, B)$ is disallowed, was not received via the routing from $P_i$, or any of the $3+m+2$ elements from offset $\overline{j}^B_i$ in his table $\overline{H}^B_i$ are used.  Discarding here does not deplete table elements.

                \item Bob retrieves and erases $\overline{R}^B_i$ (length $3 + m$) at offset $\overline{j}^B_i$ and $\overline{v}^B$ (length $2$) at offset $\overline{j}^B_i+3+m$ from the table.

                \item Bob discards the message on failure of the relation%
                \begin{aeq}
                    \overline{t}^B_i = h_{\overline{v}^B_i} (\overline{M}^B_i)	
                \end{aeq}%

                \item Bob then calculates%
                \begin{aeq}
                    Y^A_i := \overline{Z}^B + \overline{R}^B_i
                \end{aeq}%
            \end{enumerate}
        \end{enumerate}

        \item \textit{Secret reconstruction}

        \begin{enumerate}[label=(\roman*),leftmargin=0.25em]
            \item Bob assembles all sets of $k$ messages with shared $(A, B, K^A, o^A)$, keeping the associated $(x_i, Y^A_i)$.
            \item Bob solves for a candidate $Y^A_0$ in $f_{-2}(x_0) \parallel \dots \parallel f_m(x_0) = Y^A_0$ from the $(x_i, Y^A_i)$ tuples of each set, obtaining a candidate per set, and eliminates duplicates. Violated security assumptions may lead to nonduplicates.
            \item Bob partitions each distinct candidate $Y^A_0$ as:%
            \begin{aeq}
                Y^A_0 =: u^A \parallel S^A 	
            \end{aeq}%
            and forms the candidate tuple $(u^A, S^A, o^A)$. 
        \end{enumerate}
    \end{enumerate}

    \item \textbf{Secret validation}

    Bob discards each candidate tuple $(u^A, S^A, o^A)$ violating%
    \begin{aeq}
        o^A = h'_{u^A} ( S^A )
    \end{aeq}%
    Bob aborts the protocol on no remaining candidate tuples, else he terminates the protocol with the secret $S^A$ from some tuple.
    The tuple $(A, B, K^A, S^A)$ is known by both Alice and Bob.

\end{enumerate}

\begin{remark}
    Only Bob knows whether the protocol completed.
\end{remark}

}
\FloatBarrier

\bibliographystyle{IEEEtran}
\bibliography{references}

\end{document}